\documentclass[showpacs,twocolumn]{revtex4}

\usepackage{graphicx}
\usepackage{dcolumn}
\usepackage{amsmath}
\makeatletter
\def\btt#1{\texttt{\@backslashchar#1}}
\DeclareRobustCommand\bblash{\btt{\@backslashchar}}
\makeatother

\begin{document}

\title{5D Radiating black holes in Einstein-Yang-Mills-Gauss-Bonnet gravity}

\author{S. G.~Ghosh}
\email{sghosh2@jmi.ac.in; sgghosh@gmail.com} \affiliation{Centre for
Theoretical Physics, Jamia Millia Islamia, New Delhi, 110025, India}
\date{\today}

\begin{abstract}
We derive nonstatic spherically symmetric solutions of a null fluid,
in five dimension (5D), to Einstein-Yang-Mills (EYM) equations with
the coupling of Gauss-Bonnet (GB) combination of quadratic curvature
terms, namely, 5D-EYMGB radiating black hole solution.  It is shown
that, in the limit, we can recover known radiating black hole
solutions. The spherically symmetric known 5D static black hole
solutions are also retrieved.  The effect of the GB term and
Yang-Mills (YM) gauge charge on the structure and location of
horizons, of the 5D radiating black hole, is also discussed.
\end{abstract}

\pacs{04.20.Dw, 04.50.+h,14.80.Hv, 11.15.-q}

\maketitle

\section{Introduction}
 Recent years
have witnessed a renewed  interest to study black hole solutions in
string-generated gravity models which  mainly is accomplished by
studying solutions of the Einstein theory
 supplemented by Gauss-Bonnet (GB) term \cite{GB_BH,Wheeler_1}.
 String theory also predicts quantum corrections to classical gravity theory and the
GB term is the only one leading to second order differential
equations in the metric. On the
 other hand the black hole  solutions in gravity coupled to fields of different types have always
 drew in much attention,  in particular, a great interest in solutions to Einstein-Yang-Mills
 (EYM) systems \cite{py,br,cf, shmh07,shmh08,nbmd,mdab,dbp}.  Wu and Yang
 \cite{wy} obtained static symmetric solution of Yang-Mills equation for the \textit{isospin} gauge group
 $SO(3)$.  The remarkable
feature of this Wu-Yang \textit{ansatz} is that the field has no
contribution from gradient and instead has pure YM non-Abelian
component.
 A curved-space generalization of the Wu-Yang
 solutions  \cite{wy} for the gauge group  $SO(3)$ is shown to be a special case of  Yasskin's \cite{py} solutions.      It is known
that non-Abelian gauge theory coupled to gravitation, i.e., EYM
results to precisely  the geometry of Reissner-Nordstr$\ddot{o}$m
with the charge that determines the geometry is  gauge charge
\cite{py,br,cf}.  Indeed, Yasskin \cite{py} gave an explicit theorem
so that from each solution of the Einstein-Maxwell equations one can
get solutions of EYM equations.   One would like to study how these
features get modified in higher-dimensional (HD) spacetimes and
whether this theorem holds in HD spacetimes. Recent developments in
string theory indicate that gravity may
 be truly HD theory, becoming effectively  four-dimensional (4D)  at
 lower energies.  Since non-Abelian gauge fields also feature in the low energy
effective action of string theory, it is interesting to study the
properties of the corresponding  EYM in presence of GB terms.
Mazharimousavi and Halilsoy \cite{shmh07,shmh08} have found static
spherically symmetric HD black hole solutions to coupled set of
equations of the EYMGB, for $SO(N-1)$ gauge group, systems which are
based on the Wu and Yang \cite{wy} \textit{ansatz}. The
corresponding static topological black holes have been found
independently by others \cite{nbmd,mdab,dbp}.

It would be interesting to further consider nonstatic generalization
of Mazharimousavi and Halilsoy  solutions \cite{shmh07,shmh08}. It
is the purpose of this Letter to obtain an exact nonstatic solution
of the 5D EYMGB theory in  the presence of a null fluid and by
employing the Wu-Yang \textit{ansatz}. We shall present a class of
5D nonstatic solutions describing the exterior of radiating black
holes with null fluid endowed with gauge charge, i.e., an exact
Vaidya-like solution in 5D EYMGB theory. The Vaidya geometry
permitting the incorporations of the effects of null fluid offers a
more realistic background than static geometries, where all back
reaction is ignored.  It may be noted one of few nonstatic black
hole solutions is Vaidya \cite{pc} which is a solution of Einstein's
equations with spherical symmetry for a null fluid (radially
propagating radiation) source. It is possible to model the radiating
star by matching them to exterior Vaidya spacetime (see
\cite{wwgb,adsg} for reviews on Vaidya solution and \cite{sgad} for
it's higher-dimensional version). This Letter also examines the
effect of the GB terms and YM gauge charge on the structure and
location of the horizons for the radiating black holes.  A black
hole has three horizon like surface \cite{jygb,rm}: time-like limit
surface (TLS), apparent horizons (AH) and event horizons (EH). In
general the three horizon does not coincide and they are sensitive
to small perturbation.  For a classical Schwarzschild black hole
(which does not radiate), the three surfaces EH, AH, and TLS are all
identical. Upon "switching on" the Hawking evaporation this
degeneracy is partially lifted even if the spherical symmetry stays.
We have then AH=TLS,  but the EH is different from AH=TLS. In
particular, the AH is located inside the EH, the portion of
spacetime between the two surfaces forming the so-called "quantum
ergosphere". If we break spherical symmetry preserving stationarity
(e.g., Kerr black hole), then AH=EH but EH $ \neq $ TLS. Here the
ergosphere is the space between "the" horizon EH=AH and the TLS,
usually called the "static limit" \cite{jygb}. In both cases
particles and light signals can escape from within the ergosphere
and reach infinity. The characteristics of EH and AH associated with
black holes in 5D EYMGB are also discussed.
\section{Vaidya-like solution in 5D EYMGB theory}
We consider  $SO(4)$ gauge theory with structure constant $C_{\left(
\beta\right) \left( \gamma\right) }^{\left( \alpha \right) }$, the
YM fields $F_{a b }^{\left( \alpha\right) }$ and the YM potential
$A_{a }^{\left( \alpha \right)}$. The gauge potentials $A_{a
}^{\left( \alpha\right)}$ and the Yang-Mills fields $F_{a b
}^{\left( \alpha\right) }$ are related through the equation
\begin{equation}
F_{a b }^{\left( \alpha\right) }=\partial _{a }A_{b }^{\left(
\alpha\right) }-\partial _{b }A_{a }^{\left( \alpha\right)
}+\frac{1}{2\sigma }C_{\left( \beta\right) \left( \gamma\right)
}^{\left( \alpha \right) }A_{a }^{\left( \beta\right) }A_{b
}^{\left( \gamma \right) }.
\end{equation}
 We note that the internal indices $ \{\alpha,\beta,\gamma,...\}$ do not differ whether in covariant or
contravariant form. The action which describes EMYGB theory in 5D
reads \cite{shmh07,shmh08}:
\begin{equation}\label{action}
\mathcal{I_{G}}=\frac{1}{2}\int_{{M}}dx^{5}\sqrt{-g}\left[ (R +
\omega'
L_{GB})-\sum_{\alpha=1}^{6} F_{a b }^{(\alpha)}F^{(\alpha)a b }%
\right].
\end{equation}%
Here, $g$ = det($g_{ab}$) is the determinant of the metric tensor,
 $R$ is the Ricci Scalar and $\omega' = \omega/2$ with $\omega$ the coupling constant of the GB terms. This type
of action is derived in the low-energy limit of heterotic
superstring theory~\cite{Gross}. In that case, $\omega$ is regarded
as the inverse string tension and positive definite, and we consider
only the case with $\omega \ge 0$ in this paper. Expressed in terms
of Eddington advanced time coordinate (ingoing coordinate) $v$, with
the metric ansatz of 5D spherically symmetric spacetime
\cite{sgad,sgdd,gdey}:
\begin{equation}
ds^2 = - A(v,r)^2 f(v,r)\;  dv^2
 +  2 A(v,r)\; dv\; dr + r^2 d \Omega^2_3, \label{eq:me1gb}
\end{equation}
where $ d\Omega^2_3 = d \theta^2+ \sin^2 \theta d \phi^2 + \sin^2
\theta \sin^2\phi^2 d\psi^2$. Here $A$ is an arbitrary function of
$v$ and $r$ and $\{ x^a \} = \{ v,\;r,\; \theta,\; \phi,\;\psi  \}$.
  We wish to find the general
solution of the Einstein equation for the matter field given by
Eq.~(\ref{emt}) for the metric (\ref{eq:me1gb}), which contains two
arbitrary functions. It is the field equation $G^0_1 = 0$ that leads
to $ A(v,r) = g(v)$ \cite{sgad,sgdd}. This could be absorbed by
writing $d \tilde{v} = g(v) dv$.  Hence, without loss of generality,
the metric (\ref{eq:me1gb}) takes the form ,
\begin{equation}
ds^2 = - f(v,r) d v^2 + 2 d v d r + r^2 d  \Omega^2_3.
 \label{eq:megb}
\end{equation}
 We introduce the Wu-Yang \textit{ansatz} in
5D \cite{shmh07,shmh08} as
\begin{eqnarray} \label{ww}
A^{(\alpha)} &=&\frac{Q}{r^{2}}\left(
x_{i}dx_{j}-x_{j}dx_{i}\right), \;
\\
2 &\leq &i\leq 4,  \notag \\
1 &\leq &j\leq i-1 , \notag \\
1 &\leq &\left( \alpha\right) \leq 6, \notag
\end{eqnarray}%
where the super indices $\alpha$ are chosen according to the values
of $i$ and $j$ in order\cite{shmh07,shmh08}.  It is easy to see that
for the metric \ref{eq:megb}), the YM matter field equations admit
solution $\sigma = Q$ \cite{shmh07,shmh08}. The Wu-Yang solution
appears highly nonlinear because of mixing between spacetime indices
and gauge group indices.  However, it is linear as expressed in the
nonlinear gauge fields because purely magnetic gauge charge is
chosen along with position dependent gauge field transformation
\cite{py}.
The YM field 2-form is defined by the expression%
\begin{equation}
F^{\left( \alpha\right) }=dA^{\left( \alpha \right)
}+\frac{1}{2Q}C_{\left( \beta\right) \left( \gamma\right) }^{\left(
\alpha\right) }A^{\left( \beta\right) }\wedge A^{\left(
\gamma\right) }.
\end{equation}
 The integrability conditions
\begin{equation}
dF^{\left( \alpha\right) }+\frac{1}{Q}C_{\left( \beta\right) \left(
c\right) }^{\left( \alpha\right) }A^{\left( \beta\right) }\wedge
F^{\left( \gamma\right) }=0,
\end{equation}%
as well as the YM equations
\begin{equation}
d\ast F^{\left( \alpha\right) }+\frac{1}{Q}C_{\left( \beta\right)
\left( \gamma\right) }^{\left( \alpha\right) }A^{\left( \beta\right)
}\wedge \ast F^{\left( \gamma\right) }=0,
\end{equation}%
are all satisfied.  Here $d$ is exterior derivative, $\wedge$ stands
for  wedge product and $\ast$ represents Hodge duality. All these
are in the usual exterior differential forms notation. The GB
Lagrangian is of the form
\begin{equation}\label{EGB}
L_{GB} = R^2 - 4 R_{ab}R^{ab}+R_{abcd} R^{abcd}.
\end{equation}
 The action (\ref{action}) leads to the following set of field
equations:
\begin{equation}\label{FE}
\mathcal{G}_{ab} \equiv  G_{ab} + \omega' H_{ab} = T_{a b},
\end{equation}
where
\begin{eqnarray}\label{equations}
  G_{ab} &=& R_{ab} -\frac{1}{2} g_{ab} R,
\end{eqnarray}
is the Einstein tensor and
\begin{eqnarray}
  H_{ab} &=& 2[RR_{ab}-2R_{a\alpha}R^{\alpha}_b -
  2 R^{\alpha \beta}R_{a\alpha b\beta} + R_a^{\alpha\beta\gamma}
  R_{b\alpha\beta\gamma}]  \nonumber \\& & -\frac{1}{2}g_{ab}L_{GB},
\end{eqnarray}
is the Lanczos tensor. The stress-energy tensor is written as
\begin{equation}\label{emt}
T_{a b} =T_{a b }^G+T_{a b }^N,
\end{equation}
where  the gauge stress-energy tensor  $T_{a b }^G$ is
\begin{eqnarray}
T_{a b }^G=\sum_{\alpha=1}^{6} \left[ 2F_{a }^{\left( \alpha\right)
\lambda }F_{b \lambda }^{\left( \alpha\right)
}-\frac{1}{2}F_{\lambda \sigma }^{\left( \alpha\right) }F^{\left(
\alpha\right) \lambda \sigma }g_{a b }\right],
\end{eqnarray}
The energy-momentum tensor of a null fluid is
\begin{eqnarray}
{T}_{ab}=\psi(v,r) \beta_{a}\beta_{b}, \label{eq:emtgb}
\end{eqnarray}
where $\psi(v,r)$ is the non-zero energy density and $\beta_a$ is a
null vector with
\begin{eqnarray}
\beta_{a} = \delta_a^0, \beta_{a}\beta^{a} = 0.
\end{eqnarray}
Introducing
\begin{eqnarray*}
 x_1 &=& r\cos \psi \sin \phi \sin \theta, \\
 x_2 &=&  r\sin \psi \sin \phi \sin \theta, \\
 x_3 &=&  r\cos \phi \sin \theta, \\
 x_4 &=&  r\cos \theta,
\end{eqnarray*}
and using \textit{ansatz} (\ref{ww}) one obtains
\begin{eqnarray*}
  A^{(1)} &=& -Q\sin^{2} \phi \sin^{2} \theta \; d \psi,
 \\
   A^{(2)} &=& Q \sin^{2} \theta \left( \cos \psi\; d \phi -\cos \phi \sin
\psi \sin \phi \;  d \psi \right),
 \\
  A^{(3)} &=& Q  \sin^{2} \theta \left( \sin   \psi \; d \phi   + \cos \phi   \cos \psi    \sin   \phi \;  d \psi
  \right),
  \\
  A^{(4)} &=& Q \big( \sin  \theta \left( \cos  \psi \cos   \phi   d \phi   - \sin \psi  \sin \phi \; d \psi \right) \cos
  \theta \nonumber \\
   & + & \cos  \psi   \sin   \phi\; d \theta
  \big), \\
  A^{(5)} & =&  Q \left( \cos \phi\; d \theta - \cos \theta \sin \phi \sin \theta \; d \phi
  \right),
 \\
  A^{(6)} &=& Q \left( \cos \phi d \theta -\cos \theta \sin \phi \sin  \theta \; d \phi
  \right).
\end{eqnarray*}
We then find that the non-zero components would read as:
$T^r_v=\psi(v,r)$, $T_v^v = T_r^r =-3Q^2/(2r^4)$ and
$T^{\theta}_{\theta}=T^{\phi}_{\phi} = T^{\psi}_{\psi}= Q^2/r^4$.
 It may be recalled
that energy-momentum tensor (EMT) of a Type II fluid has a double
null eigenvector, whereas an EMT of a Type I fluid has only one
time-like eigenvector \cite{wwgb,hegb}.  It may be noted that, the
gauge field has only the angular components,
$F^\alpha_{\theta_i\theta_j}$ with $ i\neq j $, non-zero and they go
as $r^{-2}$ which in turn make $T_{ab}^G$  go as $r^{-4}$.

The only non-trivial components of the EGB tensor
($\mathcal{G}^a_b$), in a unit system with $\omega' = \omega /2$,
take the form:
\begin{equation}
\mathcal{G}^v_v = \mathcal{G}^r_r = f' - \frac{2}{r}(1-f)+ \frac{4\omega}{r^2}(1-f)f', \label{feqgb}\\
\end{equation}
\begin{equation}
\mathcal{G}^{\theta}_{\theta} =\mathcal{G}^{\phi}_{\phi}
=\mathcal{G}^{\psi}_{\psi} = f'' + \frac{4}{r}f' + \frac{2}{r^2}
(1-f) + \frac{4
\omega}{r^2}\left[ f''(1-f) + f{'}^2 \right], \\
\end{equation}
\begin{equation}
\mathcal{G}^r_v    = \frac{3}{2}\frac{\dot{f}}{r} + \frac{6
\omega}{r^3}\dot{f}(1-f).\label{feqgb1}
\end{equation}
Then, $f(v,r)$ is obtained by solving only the  (\ref{FE}),  the
equation $\mathcal{G}^v_v = T^v_v$ is integrated to give the general
solution as
\begin{equation}
f(v,r) = 1 + \frac{r^2}{2 \omega} \left[ 1 \pm \sqrt{1 + \frac{4
\omega M(v)}{r^4} - \frac{8 \omega Q^2 \ln r}{r^4}+\frac{4
\omega^2}{r^4}}\right], \label{s-feqgb}
\end{equation}
where $M(v)$ is positive and an arbitrary function of $v$ identified
as mass of the matter. The gauge charge $Q$ can be either positive
or negative. The special case in which $\dot{M}(v) = 0 $ and $Q^2 =
0$, Eq.~(\ref{feqgb}) leads to GB-Schwarzschild solution, of which
the global structure is presented in~\cite{tmgb}. The solution
(\ref{s-feqgb}) is a general spherically symmetric solution of the
5D EYMGB theory with the metric (\ref{eq:megb}) for the null fluid
defined by the energy-momentum tensor (\ref{eq:emtgb}). Since YM
$T_{ab}^G$ go as $r^{-4}$ (the same as for Maxwell field in $D=4$),
for $5D$. That is why its contribution in $f(v,r)$ will be the same
for $5D$ as in 4D Reissner-Nordstr$\ddot{o}$m (RN)  black hole
\cite{gdey}. The nonradiating limit of this would be 5D-Yaskin black
hole and not 5D analogue of Reissner-Nordstr$\ddot{o}$m.

 There are two families of solutions which
correspond to the sign in front of the square root in
Eq.~(\ref{s-feqgb}). We call the family which has the minus (plus)
sign the minus- (plus+) branch solution. From $\mathcal{G}^r_v =
T^r_v$, we obtain the energy density of the null fluid as
\begin{eqnarray}
\psi(v,r) =\frac{3}{2}\frac{\dot{M(v)}}{r^3}. \label{density}
\end{eqnarray}
for both branches, where the dot denotes the derivative with respect
to $v$. We first turn our attention to the three limiting cases when
the solution is known. These are (i) $M(v) \neq 0,  \; Q =0$ and
$\omega \neq 0$ then 5D-EGB black holes \cite{tk,hm,sgdd}.  The
solution of the Eq.~(\ref{feqgb}) is
\begin{equation}\label{egb}
f(v,r) = 1 + \frac{r^2}{2 \omega} \left[ 1 \pm \sqrt{1 + \frac{4
\omega M(v)}{r^4}} \right],
\end{equation}
(ii) $M(v) \neq 0,  \; Q \neq 0 $ and $\omega  = 0$ then the 5D-EYM
black holes \cite{gdey}.  Now one has solution of the
Eq.~(\ref{feqgb}) as
\begin{eqnarray}
f(v,r) =1-\frac{M(v)}{r^{2}} - \frac{2Q^2 \ln r}{r^2},
\label{vaidyaEYM}
\end{eqnarray}
 and (iii) in the general relativistic limit ${ \omega} \to 0$ and $Q^2 \to
0$, the minus-branch solution reduces to
\begin{eqnarray}
f(v,r) =1-\frac{M(v)}{r^{2}}, \label{vaidyaGR}
\end{eqnarray}
which is the $5D$ Vaidya solution \cite{sgad,sgdd} in Einstein
theory. It may be noted that, in $5D$ Einstein theory, the density
is still given by Eq.~(\ref{density}). There is no such limit for
the plus-branch solution.  The family of solutions discussed here
belongs to Type II fluid. However, In the static case  $M = $
constant and the matter field degenerates to Type I fluid, we can
generate static black hole solutions obtained in
\cite{shmh07,shmh08} by proper choice of these constants. In the
static limit, this metric can be obtained from the metric in the
usual spherically symmetric form,
\begin{equation}
ds^2 = -f(r)\; dt^2 + \frac{dr^2}{f(r)} + r^2 (d \Omega_{3})^2.
\end{equation}
with
\begin{equation}
f(r) = 1 + \frac{r^2}{2 \omega} \left[ 1 \pm \sqrt{1 + \frac{4
\omega M}{r^4} - \frac{8 \omega Q^2 \ln r}{r^4}+\frac{4
\omega^2}{r^4}}\right], \label{static}
\end{equation}
if $Q^2 \to 0$ this solution reduces to the solution which was
independently discovered by Boulware and Deser~\cite{GB_BH} and
Wheeler~\cite{Wheeler_1}.

 The Kretschmann scalar ($K = R_{abcd}
R^{abcd}$, $R_{abcd}$ is the 5D Riemann tensor ) and Ricci scalar
($R = R_{ab} R^{ab}$, $R_{ab}$ is the 5D Ricci tensor) for the
metric (\ref{eq:megb}) reduces to
\begin{equation}
\mbox{K} = f{''}^2+\frac{6}{r^4}f{'}^2 + \frac{12}{r^4}
(1-f)^2,\label{ksgb}
\end{equation}
and
\begin{equation}
R = f{''}+ \frac{6}{r}f'- \frac{6}{r^2} (1-f),\label{rsgb}
\end{equation}
Radial ($ \theta$ and $
\phi \,=\,const$.) null geodesics of the metric (\ref{eq:megb}) must
satisfy the null condition
\begin{equation}
2 \frac{dr}{dv} =1 + \frac{r^2}{4 \omega} \left[ 1 \pm \sqrt{1 +
\frac{4 \omega M(v)}{r^4} - \frac{8 \omega Q^2 \ln r}{r^4}+\frac{4
\omega^2}{r^4}}\right], \label{eq:de1gb}
\end{equation}
The invariants are regular everywhere except at the origin $r = 0$,
where they diverge.  Hence, the spacetime has the scalar polynomial
singularity \cite{hegb} at $r=0$.  The nature (a naked singularity
or a black hole) of the singularity can be characterized by the
existence of radial null geodesics emerging from the singularity.
The singularity is at least locally naked if there exist such
geodesics, and if no such geodesics exist, it is a black hole. The
study of causal structure of the spacetime is beyond the scope of
this paper and will be discussed elsewhere \cite{pg}.

\noindent \emph{{Energy conditions}}: The family of solutions
discussed here, in general, belongs to Type II fluid defined in
\cite{hegb}. When $m=m(r)$, we have $\psi$=0, and the matter field
degenerates to type I fluid \cite{wwgb,adsg}. In the rest frame
associated with the observer, the energy density of the matter will
be given by
\begin{equation}
\mu = T^r_v,\hspace{.1in}\rho = - T^t_t = - T^r_r  \label{energy}
\end{equation}
 and the principal pressures are $P_i = T^i_i$ (no sum convention) and due to isotropy  $P=P_i$ for all $i$.
\noindent \emph{a) The weak energy conditions} (WEC): The
energy-momentum tensor obeys inequality $T_{ab}w^a w^b \geq 0$ for
any time-like vector \cite{hegb}, i.e., $\psi \geq 0,\hspace{0.1
in}\rho \geq 0,\hspace{0.1 in} P   \geq 0,$. We say that strong
energy condition (SEC), holds for Type II fluid if, WEC is true.,
i.e., both WEC and SEC,
for a Type II fluid, are identical \cite{wwgb}. \\
\noindent {\emph{b) The dominant energy conditions (DEC) }}: For any
time-like vector $w_a$, $T^{ab}w_a w_b \geq 0$, and $T^{ab}w_a$ is
non-space-like vector , i.e., $ \psi \geq 0,\hspace{0.1 in}\rho \geq
P \geq 0.$  Hence WEC and SEC are satisfied if $\dot{M}(v) \geq 0$.
In addition DEC also holds.
\section{Horizons of 5D radiating black hole}
 The line element of the radiating black hole in
5D EYMGB theory has the form (\ref{eq:megb}) with $f(v,r)$ given by
Eq.~(\ref{s-feqgb}) and  the energy-momentum tensor
(\ref{eq:emtgb}). The luminosity due to loss of mass is given by $L
= - dM/dv$,  where $L < 1$.  Both  are measured in the region where
$d/dv$ is time-like. In order to further discuss the physical nature
of our solutions, we introduce their kinematical parameters. As
first demonstrated by York \cite{jygb} and later by others
\cite{rm,sgdd}, the horizons may be obtained to $O(L)$ by noting
that a null-vector decomposition of the metric (\ref{eq:megb}) is
made of the form
\begin{equation}\label{gabgb}
g_{ab} = - \beta_a l_b - l_a \beta_b + \gamma_{ab},
\end{equation}
where,
\begin{eqnarray}
\beta_{a} &=& - \delta_a^v, \: l_{a} = - \frac{1}{2} f(v,r)
\delta_{a}^v + \delta_a^r, \label{nvagb}
 \\
\gamma_{ab} &=& r^2 \delta_a^{\theta} \delta_b^{\theta} + r^2
\sin^2(\theta) \delta_a^{\varphi} \delta_b^{\varphi}
 \nonumber\\&&+ r^2 \sin^2(\theta)\sin^2(\phi) \delta_a^{\psi} \delta_b^{\psi},
\label{nvbgb}
\\
l_{a}l^{a} &=& \beta_{a} \beta^{a} = 0, \; ~l_a \beta^a = -1,\; ~l^a
\;\gamma_{ab} = 0, \nonumber\\&& ~\gamma_{ab}\; ~\beta^{b} = 0,
\label{nvdgb}
\end{eqnarray}
 with $f(v, r)$ given by Eq.~(\ref{s-feqgb}).  The Raychaudhuri
equation of null-geodesic congruence is
\begin{equation}\label{regb}
   \frac{d \Theta}{d v} = \kappa \Theta - R_{ab}l^al^b-\frac{1}{2}
   \Theta^2 - \sigma_{ab} \sigma^{ab} + \Omega_{ab}\Omega^{ab},
\end{equation}
with expansion $\Theta$, twist $\Omega$, shear $\sigma$, and surface
gravity $\kappa$. The expansion of the null rays parameterized by
$v$ is given by
\begin{equation}\label{theta}
\Theta = \nabla_a l^a - \kappa,
\end{equation}
where the $\nabla$ is the covariant derivative and the surface
gravity is
\begin{equation}\label{sggb}
\kappa = - \beta^a l^b \nabla_b l_a.
\end{equation}
In the case of spherical symmetric, the vorticity and shear of $l_a$
are zero. Substituting Eqs.~(\ref{s-feqgb}) and (\ref{nvagb}) into
(\ref{sggb}), we obtain surface gravity
\begin{eqnarray}
\kappa & = & \frac{r}{2 \omega} \left[1 - \sqrt{1 + \frac{4 \omega
M(v)}{r^4} - \frac{8 \omega Q^2 \ln r}{r^4}+\frac{4
\omega^2}{r^4}}\right] \\ \nonumber & &  + \frac{\frac{2M(v)}{r^3} +
\frac{Q^2}{r^3}-\frac{4Q^2 \ln r}{r^3}+\frac{2 \omega}{r^3}}{\sqrt{1
+ \frac{4 \omega M(v)}{r^4} - \frac{8 \omega Q^2 \ln r}{r^4}+\frac{4
\omega^2}{r^4}}}.\label{Kgb}
\end{eqnarray}
Then Eqs.~(\ref{s-feqgb}), (\ref{nvagb}), (\ref{theta}),  and
(\ref{Kgb}) yield the expansion of null ray congruence:
\begin{equation}
\Theta = \frac{3}{2r} \left[1 + \frac{r^2}{4\omega}\left[1 - \sqrt{1
+ \frac{4 \omega M(v)}{r^4} - \frac{8 \omega Q^2 \ln r}{r^4}+\frac{4
\omega^2}{r^4}}\right]\right]. \label{thgb}
\end{equation}
The apparent horizon (AH) is the outermost marginally trapped
surface for the outgoing photons.  The AH can be either null or
space-like, that is, it can 'move' causally or acausally
\cite{jygb}. The apparent horizons are defined as surface such that
$\Theta \simeq 0$ which implies that $f=0$.  From the
Eq.~(\ref{thgb}) it is clear that AH is the solution of
\begin{equation}
\left[1 + \frac{r^2}{2\omega}\left[1 - \sqrt{1 + \frac{4 \omega
M(v)}{r^4} - \frac{8 \omega Q^2 \ln r}{r^4}+\frac{4
\omega^2}{r^4}}\right]\right] = 0.
\end{equation}
i.e.,  zeros of
\begin{equation}\label{aheq}
r^2 - M(v) + 2 Q^2 \ln(r) = 0.
\end{equation}
For $Q \rightarrow 0$ and constant $M$, we have 5D Schwarzschild
horizon $r=\sqrt{M}$.  In general, Eq.~(\ref{aheq}), which admits
solutions
\begin{equation}\label{ahym}
    r_{IAH} = \exp \left[ -\frac{1}{2} \frac{Q^2 \mbox{LambertW} \left(0, x\right)+M(v)}{Q^2}\right].
\end{equation}
\begin{equation}\label{ahym}
    r_{OAH} = \exp \left[ -\frac{1}{2} \frac{Q^2 \mbox{LambertW} \left(-1, x\right)+M(v)}{Q^2}\right].
\end{equation}
Here $ x= - \frac{\exp (-M(v)/Q^2)}{Q^2}.$

Here $r_{IAH}$ and $r_{OAH}$ are respectively inner and outer
horizons and the LambertW function satisfies
\[\mbox{LambertW}(x)\exp \left[\mbox{LambertW}(x)\right]=x.\] The
important feature of Eq.~(\ref{ahym}) is that it is $\omega$
independent. This lead to fact that it is similar to pure EYM case.
Thus the GB term does not cause the AHs of the 5D EYM black holes to
be distorted. The TLS for a black hole, with a small luminosity, is
locus where $g_{vv} = 0$. Here one sees that $\Theta = 0$, implies
$f=0$ or $g_{vv}(r=r_{AH}) = 0$ implies that $r=r_{AH}$ is  TLS and
AH and TLS coincide in our non-rotational case. The pure charged
case ($M(v)$ = 0) is also important, then we have horizon without
mass
\begin{equation}\label{ahq}
r_{IAH} = \exp \left[-\frac{1}{2}\mbox{LambertW}\left(
0,\frac{1}{Q^2}\right) \right],
\end{equation}

\begin{equation}\label{ahq}
r_{OAH} = \exp \left[-\frac{1}{2}\mbox{LambertW}\left(
-1,\frac{1}{Q^2}\right) \right].
\end{equation}
For an outgoing null geodesic, $\dot{r}$ is given by
Eq.~(\ref{eq:de1gb}). Differentiation of (\ref{eq:de1gb}) w.r.t. $v$
gives
\begin{eqnarray}\label{rdd}
\ddot{r} & = & \frac{r\dot{r}}{2\omega}\left[1 - \sqrt{1 + \frac{4
\omega M(v)}{r^4} - \frac{8 \omega Q^2 \ln r}{r^4}+\frac{4
\omega^2}{r^4}}\right]  \\ & & + \nonumber \frac{\frac{L}{2r^2} +
\frac{2 M(v)\dot{r}}{r^3} + \frac{Q^2 \dot{r}}{r^3}- \frac{4 Q^2 \ln
r \dot{r}}{r^3}+\frac{2 \omega \dot{r}}{r^3}}{\sqrt{1 + \frac{4
\omega M(v)}{r^4} - \frac{8 \omega Q^2 \ln r}{r^4}+\frac{4
\omega^2}{r^4}}}.
\end{eqnarray}
At the time-like surface $r = r_{AH}$, $\dot{r} = 0$ and $\ddot{r}>
0$ for $L > 0$.  Hence photons escape from $r=r_{AH}$ to the to
reach arbitrarily large distances from the hole.

However, In the general GB term does change the location of AH,
e.g., in the limit $Q  \rightarrow 0$ , in the 5D-EGB case the AHs
reads   \cite{sgdd}
\begin{equation}\label{aegb}
r_{AH} = \sqrt{M(v) - 2\omega}
\end{equation}

 Further, In the relativistic limit $\omega
\rightarrow 0, Q \rightarrow 0$ then $r_{AH} \rightarrow
\sqrt{M(v)}$.

The future event horizon (EH) is the boundary of the causal past of
future null infinity, and it represents the locus of outgoing
future-directed null geodesic rays that never manage to reach
arbitrarily large distances from the hole.  This definition requires
knowledge of the entire future history of the hole. However, York
\cite{jygb}, for a radiating black holes, argued that the question
of the escape versus trapping of null rays is, physically, matter of
qualitative degree and proposed a working definition of definition
as follows: the EH are strictly null and are defined to order of
$O(L)$ and Photons are in captivity by event horizon for times long
compared to dynamical scale of the hole.
 It can be seen to be equivalent to the
requirement that for acceleration of null-geodesic congruence at the
EH,
\begin{equation}
\left[\frac{d^2r}{dv^2}\right]_{\mbox{EH}} \simeq ~ 0. \label{ehgb}
\end{equation}
This criterion enables us to distinguish the AH and the EH to
necessary accuracy. An outgoing radial null geodesic which is
parameterized by $v$ satisfies
\begin{equation}
\frac{dr}{dv} = \frac{1}{2}\left[1 + \frac{r^2}{4\omega}\left[1 -
\sqrt{1 + \frac{4 \omega M(v)}{r^4} - \frac{8 \omega Q^2 \ln
r}{r^4}+\frac{4 \omega^2}{r^4}}\right]\right]. \label{rnggb}
\end{equation}
Then Eqs.~(\ref{Kgb}) and (\ref{thgb}) can be used to put
Eq.~(\ref{ehgb}) in the form
\begin{eqnarray}
\kappa \Theta_{EH} && \simeq  \left[ \frac{3}{2r}\frac{\partial
f}{\partial v} \right]_{EH} \nonumber \\\ && \simeq
\frac{1}{2r_{EH}^3} \frac{3L }{\sqrt{1 + \frac{4 \omega
M(v)}{r_{EH}^4} - \frac{8 \omega Q^2 \ln r_{EH}}{r_{EH}^4}+\frac{4
\omega^2}{r_{EH}^4}}}, \label{eh1gb}
\end{eqnarray}
where the expansion is
\begin{eqnarray}
\Theta_{EH}\simeq \frac{3}{2 r_{EH}}\Big[1 +
\frac{r^2_{EH}}{4\omega}\Big[1 - \nonumber \\ \sqrt{1 + \frac{4
\omega M(v)}{r_{EH}^4} - \frac{8 \omega Q^2 \ln
r_{EH}}{r_{EH}^4}+\frac{4 \omega^2}{r_{EH}^4}}\Big]\Big].
\end{eqnarray}
For the null vectors $l_a$ in Eq.~(\ref{nvagb}) and the component of
energy-momentum tensor yields
\begin{equation}
R_{a b}l^{a}l^{b} =  \frac{3}{2r} \frac{\partial f}{\partial v}.
\label{chigb}
\end{equation}
The Raychaudhuri equation, with $\sigma = \Omega = 0$ \cite{jygb}:
\begin{equation}
   \frac{d \Theta}{d v} = \kappa \Theta - R_{ab}l^al^b- \frac{1}{2}
   \Theta^2.  \label{mregb}
\end{equation}
Since EH are defined to $O(L)$, we neglect $\Theta^2$, as $\Theta^2
= O(L^2)$. Eqs.~(\ref{eh1gb}), (\ref{chigb}) and (\ref{mregb}),
imply that
\begin{equation}
  \left[ \frac{d \Theta}{d v} \right]_{EH} \simeq 0.\label{eh}
\end{equation}
Following \cite{jygb,sgdd}, for low luminosity, the surface gravity
$\kappa$ can be evaluated at $AH$ and the expression for the EH can
be obtained to $O(L)$.  It can be shown that the expression for the
event horizon is the same as that for the apparent horizon with $M$
being  replaced by $M^*$ \cite{sgdd}, where $M^*$  is effective mass
 defined as follows: $M^{*}(v) = M(v) - {L}/{\kappa}$.
From the Eq.~(\ref{aegb}), it is clear that, in general, the
presence of the coupling constant, of the GB terms, $\omega$
produces a change in the location of horizons. Such a change could
have a significant effect in the dynamical evolution of these
horizons. For nonzero $\omega$ the structure of the horizons is
non-trivial. However, the eq.~(\ref{ahym}) is independent of the
 of the GB coupling constant $\omega$, i.e., AH are exactly same as
 that in EYM without GB coupling constant  $\omega$. Thus the GB term does not alter the horizons of the
5D EYM black holes.

\section{Discussion and Conclusion}
  In this Letter we have obtain an
exact black hole solution that describes a null fluid in the
framework of 5D EYMGB theory by employing 5D curved-space
generalization Wu-Yang \textit{ansatz}.  Thus we have an explicit
nonstatic radiating black hole solution of 5D EYMGB theory. We have
used the solution to discuss the consequence of GB term and YM
charge on the structure and location of the horizons 5D radiating
black hole. The AHs are obtained exactly and EHs are obtained to
first order in luminosity by method developed by York \cite{jygb}.
We shown that a 5D radiating black hole in EYGB has three important
horizon-like loci that full characterizes its structure, viz. AH, EH
and TLS and we have relationship of the three surfaces $r_{EH} <
r_{AH} = r_{TLS}$ and the region between the AH and EH is defined as
\emph{quantum ergosphere.} The presence of the coupling constant of
the Gauss-Bonnet terms $\omega$ produces a change in the location of
these horizons \cite{sgdd}. Such a change could have a significant
effect in the dynamical evolution of these horizons.  However, It
turns out that the presence of the coupling constant of the GB terms
$\omega > 0$ does not alter the location of the horizons from the
analogous EYM case, i.e., horizons of the 5D EYM and 5D EYMGB are
absolutely same when obtained by procedure suggested by York
\cite{jygb} to $O(L)$ by a null-vector decomposition of the metric.

In 4D, the Vaidya-like solution of EYM yields the same results as
one would expect for the charged null fluid in EM theory, i.e., the
geometry is precisely of the charged-Vaidya form and the charge that
determines the geometry is magnetic charge.  This is because
$T^G_{ab}$ go over $r^{-4}$ which is exactly the same as energy
momentum of EM theory.  However, this does not hold in 5D case
because components of energy-momentum tensor for EM and EYM theories
are not same. Thus the Yaaskin's \cite{py} theorem does not hold in
5D case. The 5D solution discussed here incorporates a logarithmic
term unprecedent in 4D.

The family of solutions discussed here belongs to Type II fluid.
However, if $M =$ constant and the matter field degenerates to Type
I fluid, we can generate static black hole solutions obtained in
\cite{shmh07,shmh08} by proper choice of these constants. In
particular, our results in the limit $\omega \rightarrow 0$ and $Q
\rightarrow 0$ reduce \emph{vis-$\grave{a}$-vis} to 5D relativistic
case.

\section*{Acknowledgments}
We are grateful to the referee for a number of helpful suggestions
for improvement in the Letter. The work is supported by  university
Grants Commission (UGC) major research project grant F. NO.
39-459/2010 (SR). The author also thanks IUCAA for hospitality,
while part of this work was being done.

\end{document}